\documentclass[twocolumn]{aastex631}

\usepackage{comment}
\begin{document}

\title{Effect of Spherical Polarization on the Magnetic Spectrum of the Solar Wind}

\author{Corina Dunn}
\affiliation{Space Sciences Laboratory, University of California, Berkeley, CA 94720-7450, USA}
\affiliation{Physics Department, University of California, Berkeley, CA 94720-7300, USA}
\correspondingauthor{Corina Dunn}
\email{cidunn@berkeley.edu}

\author{Trevor Bowen}
\affiliation{Space Sciences Laboratory, University of California, Berkeley, CA 94720-7450, USA}

\author{Alfred Mallet}
\affiliation{Space Sciences Laboratory, University of California, Berkeley, CA 94720-7450, USA}

\author{Samuel Badman}
\affiliation{Center for Astrophysics $\mid$ Harvard  $\mathbin{\&}$ Smithsonian, Cambridge, MA 02138, USA}

\author{Stuart D. Bale}
\affil{Physics Department, University of California, Berkeley, CA 94720-7300, USA}
\affil{Space Sciences Laboratory, University of California, Berkeley, CA 94720-7450, USA}



\begin{abstract}
 Magnetic fluctuations in the solar wind are often observed to maintain constant magnitude of the magnetic field in a manner consistent with spherically-polarized large-amplitude Alfvén waves. We investigate the effect of spherical polarization on the magnetic spectral index through a statistical survey of magnetic fluctuations observed by Parker Solar Probe between 20$R_\odot$ and 200$R_\odot$. We find that deviations from spherical polarization, i.e., changes in $|\mathbf{B}|$ (compressive fluctuations) and one-dimensional discontinuities, have a dramatic effect on the scaling behavior of the turbulent fluctuations. We show that shallow $k^{-3/2}$ spectra are only observed for constant magnetic field strength, three-dimensional structures, which we identify as large amplitude Alfvén waves. The presence of compressive fluctuations coincides with a steepening of the spectrum up to $k^{-5/3}$. Steeper power law scalings approaching $k^{-2}$ are observed when the fluctuations are dominated by discontinuities. Near-sun fluctuations are found to be the most spherically polarized, suggesting that this spherical state is fundamental to the generation of the solar wind. With increasing distance from the Sun, fluctuations are found to become less three dimensional and more compressive, which may indicate the breakdown of the Alfvénic equilibrium state.

\end{abstract}



\section{Introduction} \label{sec:intro}
     The solar wind is a collisionless magnetized plasma characterized by nonlinear turbulent interactions through which energy cascades from large to small scales \citep{bruno2013solar}. The energy spectra of solar wind fluctuations typically follow power-law type distributions; the power law index, $\gamma$, of these fluctuations is an important diagnostic for our understanding of the relevant nonlinear processes that cascade energy from large to small scales. Equivalently, the scale dependence of the fluctuation amplitudes can be described by the second-order structure function scaling exponent, $\alpha_B$, with $\alpha_B = -(1+\gamma)$ for an asymptotically long inertial range. Observations from the Parker Solar Probe mission show that the scaling of turbulent magnetic field fluctuations in the inner heliosphere is  $\alpha_B = 1/2$ \citep{chen2020evolution, sioulas2023magnetic}, which is consistent with three-dimensional, anisotropic turbulence \citep{chandran2015intermittency, mallet2017statistical, Boldyrev2006Spectrum}. The $\alpha_B = 1/2$ scaling has often been recovered by simulations \citep{mallet2016measures, perez2009role, chandran2019reflection, meyrand2019fluidization}; however, magnetic field fluctuations at 1AU typically have a steeper scaling with $\alpha_B = 2/3$ \citep{podesta2010kinetic, chen2013residual, wicks2013correlations, roberts2010evolution, horbury2008anisotropic}.
     
    The mode composition of the fluctuations may contribute to the observed spectral index \citep{sioulas2023magnetic, bowen2018impact, podesta2010scale, chen2013residual}. The observed fluctuations in the solar wind exhibit characteristics consistent with large amplitude Alfvén waves, such as high cross helicity and constant magnetic field magnitude \citep{de2020switchbacks, bale2019highly, chen2020evolution, mcmanus2020cross, chaston2020mhd, belcher1969large, goldstein1974theory, lichtenstein1980dynamic}. This constant magnitude condition ($|\mathbf{B}| = const.$) can be described as spherical polarization, in which the magnetic field vector rotates on the surface of a sphere with radius $|\mathbf{B}|$. Measurements of the velocity fluctuations in the solar wind also show the signature of spherical polarization, further indicating the presence of large amplitude Alfvén waves \citep{wang2012large, matteini2015ion, raouafi2023parker}.While strong spherical polarization is present, there are persistent subdominant fluctuations in $|\mathbf{B}|$. These compressive fluctuations are relatively poorly understood, and have been attributed to a variety of sources. The magnetosonic slow mode, with small contributions from the fast mode may make up the composition of the compressible fluctuations \citep{howes2012slow, verscharen2017kinetic,klein2012using, chaston2020mhd}. Pressure balance structures, the perpendicular limit of the slow mode, have also been suggested \citep{tu1995mhd, yao2011multi, yao2013small}. The slow and fast waves may be subject to strong collisionless damping \citep{barnes1966collisionless}, but may be continually produced through various methods, e.g. the parametric decay instability \citep{tenerani2013parametric, derby1978modulational, jayanti1993parametric}, or shearing\citep{roberts1992velocity}; the damping may also be suppressed in the presence of background turbulence \citep{meyrand2019fluidization}. \cite{chapman2007quantifying} suggest that the steepened spectrum at 1AU can be explained by independent scaling for the parallel (compressive) and perpendicular component of the fluctuations, where the compressive component scales with $\alpha_B = 2/3$ and the perpendicular component scales with $\alpha_B = 1/2$. 
    
    Discontinuities in the solar wind can also cause a deviation from the Alfvénic spherically polarized state \citep{bruno2001identifying}. One dimensional discontinuities typically have $\alpha_B = 1$ scaling \citep{li2011effect, borovsky2010contribution}, and can affect the spectral index of the solar wind: the 2/3 scaling observed at 1AU has been attributed to their presence \citep{li2011effect, borovsky2010contribution}. Intermittency in the turbulent spectra has also been attributed to discontinues, and numerical methods of removing discontinuities have recovered $\alpha_B = 1/2$ scaling \citep{salem2009solar}. Discontinuities in the solar wind have been found to be mostly rotational, with some tangential discontinuities \citep{neugebauer2006comment}. Tangential discontinuities admit no plasma flow, while rotational discontinuities are characterized by a large deflection of the magnetic field with no change in magnitude. Tangential discontinuities may be plasma barriers between two flux tubes \citep{bruno2001identifying}. Alternatively, they are the zero-width limit of pressure balance structures, which are non-propagating structures characterized by a constant total pressure \citep{tu1995mhd}. Rotational discontinuities are typically thought to be steepened Alfvén waves \citep{neugebauer2006comment}. The magnetic ``switchback" boundaries observed by Parker Solar Probe have also been analyzed as discontinuities, with similar distributions of discontinuity type as found in the 1AU solar wind \citep{larosa2021switchbacks,akhavan2021discontinuity}.

    This work comprises a statistical survey of the magnetic field's fluctuation geometry in order to examine the effects of deviation from the spherically-polarized state on the scaling behavior. We find that steepening from $\alpha_B \approx 1/2$ to $\alpha_B \approx 2/3$ can be attributed to the presence of compressive fluctuations, while steeper structure functions up to $\alpha_B \approx 1$ are observed when fluctuations are dominated by discontinuities. The degree of spherical polarization depends on the solar distance, with fluctuations becoming less three dimensional and more compressive. The spherical state of those observations closest to the Sun suggests that this state is fundamental to the origins of the wind.

\section{Methods} \label{sec:methods}

\begin{figure*}[ht!]
    \plotone{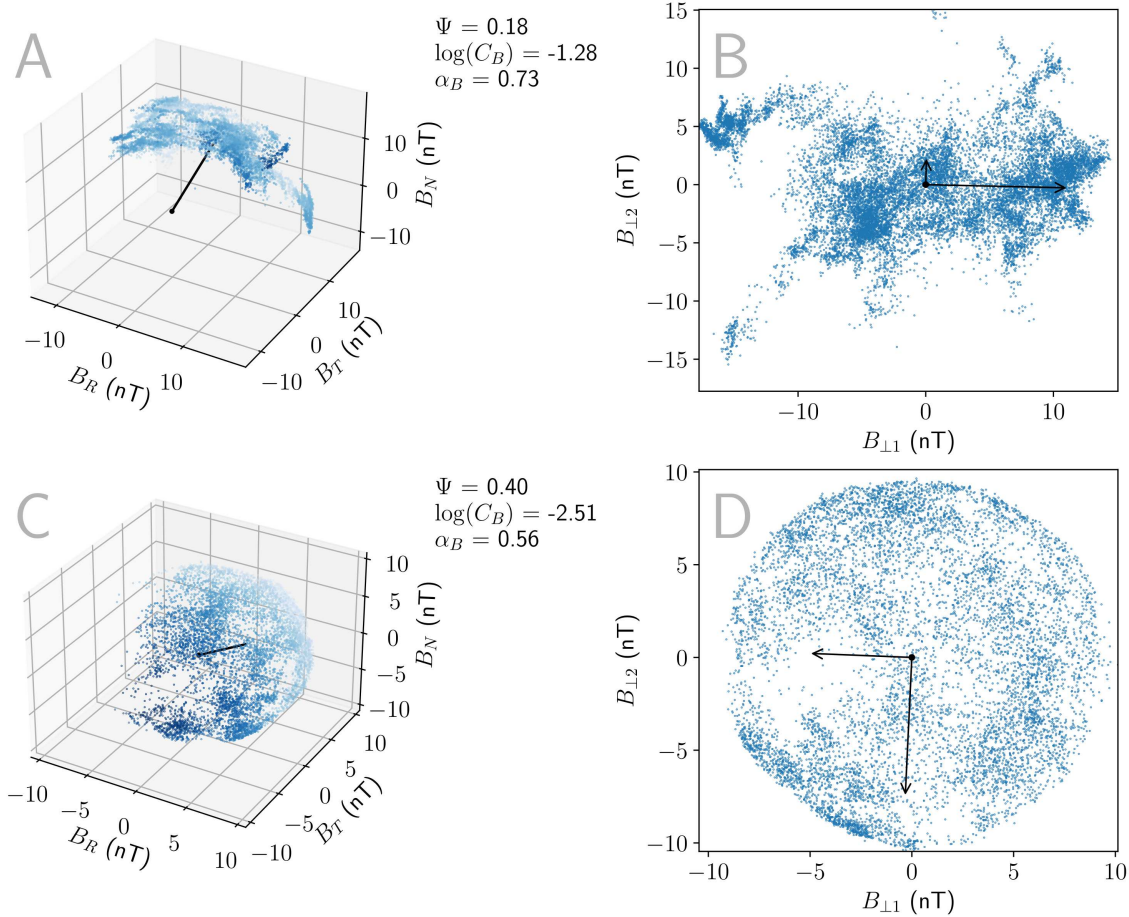}
    \caption{Panels A and C show hodograms of the magnetic field for a compressible arc-like interval (A) and for an incompressible spherically polarized interval. The mean field vector is shown as a black line from the origin. Panel B shows the perpendicular magnetic field of the structure in panel A, with its maximum and minimum variance directions shown as arrows scaled by the size of the corresponding eigenvalue. Panel D likewise shows the perpendicular magnetic field of panel C.
    \label{fig:hodograms}}
\end{figure*}

    \begin{figure*}[ht!]
    \plotone{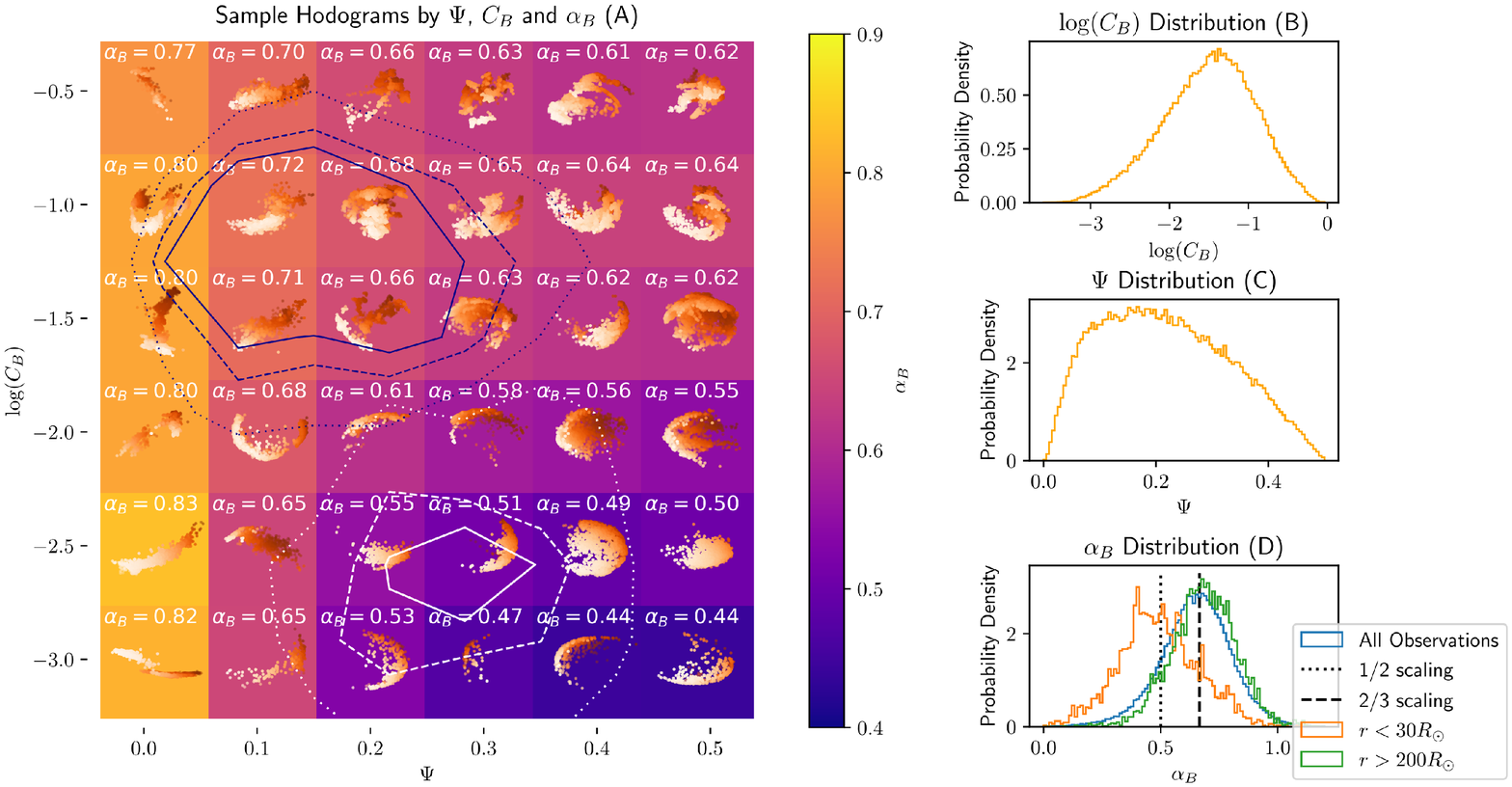}
    \caption{ \textit{Panel A:} The perpendicular variance isotropy, $\Psi$, along the x-axis, and compressibility, $\log(C_B)$, along the y-axis, form a basis to describe the geometry of observed structures. Each depicted structure is the hodogram of an interval of the magnetic field chosen to be as close as possible in $\Psi$ and $\log(C_B)$ values to its marked position. Hodograms are color-coded by the distance into the page, with the lightest points closest to the viewer. Each square is color-coded by the average $\alpha_B$ in the $\Psi$-$C_B$ space contained. This $\alpha_B$ value is also recorded in white in the top of each square. The white and purple curves respectively show contours for observations inside of 30 $R_\odot$ and outside of 200 $R_\odot$. The solid, dashed, and dotted lines show levels along which there are respectively 3.5$\%$, 3$\%$, and 2$\%$ of counts in a 100 bin 2D histogram. Panels B and C respectively show the probability density of $\log(C_B)$ and $\Psi$ with 100 bins. Similarly, panel D shows the distribution of $\alpha_B$, in blue, including 1/2 and 2/3 scalings as dotted and dashed lines respectively. The distribution of $\alpha_B$ when only the data within 30$R_\odot$ is shown in orange, with the distribution of $\alpha_B$ outside of 200$R_\odot$ shown in green. 
    \label{fig:3dminiatures}}
    \end{figure*}
    
Magnetometer time-series data from the Parker Solar Probe (PSP) \citep{fox2016solar} Fields Experiment (FIELDS) Encounters 1 through 8 are used, with a 1 second cadence \citep{bale2016fields}. Location data for the spacecraft are from the JPL Horizons database \citep{giorgini2015status}. Solar wind velocity data are unavailable or unreliable in many of the studied intervals, and was thus excluded. We plan to analyze a reliable subset of these measurements in a future study.

Data are sorted into intervals that start every 500 seconds; the duration of the interval is set to equal 10 correlation times. Correlation times, $T_C(t)$, where $t$ is the position in the timeseries of the start of the interval, are determined as the time it takes for the autocorrelation function, $C(t, \tau)$ to drop to $1/e$. In particular,
\[            C(\tau) =  \frac{\sum \delta \mathbf{B}(t) \cdot \delta \mathbf{B}(t+\tau)}{\sum |\delta \mathbf{B}|^2}
\]\[
        T_C(t) = \min\{\tau : C(t, \tau) < 1/e\},
\]
where sums are over 8000 seconds, and $\delta$ represents the deviation from the mean -- i.e. $\delta \mathbf{B}(t) = \mathbf{B}(t)-\langle\mathbf{B}\rangle$. If the correlation time was greater than 5000 seconds, the interval was discarded, requiring a maximum interval size of 50000 seconds. Intervals were on average 7092 seconds. Additionally, intervals where more than $5\%$ of data was missing or where the magnitude squared of the magnetic field is less than 5nT on average were discarded, so that observed structures are resolvable within instrument precision. Any remaining missing data are ignored. In this way, a total of 101117 intervals are collected over the studied encounters. Results were affected minimally when non-overlapping intervals are used -- the significant overlap is chosen so that detailed statistics can be obtained at every studied solar distance. Once the correlation times are computed, we compute several parameters over each interval.

We define the compressibility of the magnetic field, $C_B$ \citep{chen2020evolution} as a ratio squared of the compressible fluctuations to the incompressible fluctuations, in particular
\begin{equation}\label{eq:CB}
    C_B = \frac{\langle\delta|\mathbf{B}|^2\rangle}{\langle|\delta \mathbf{B}|^2\rangle}  = \frac{\sigma_{|\mathbf{B}|}^2}{tr(\mathbf{K}_{\mathbf{B}})},
\end{equation}
where $\mathbf{K_{\mathbf{B}}}$ is the covariance matrix of the 3-component magnetic field, and angle brackets are an average over the studied interval.

To distinguish between two and three dimensional structures, we develop a measure of the ``three dimensionality" of observations. We use a method similar to that employed by \cite{bruno2001identifying}, in which the power of the variation along the minimum and maximum variance directions is compared, but we consider only the perpendicular fluctuations because the constraint to the surface of a sphere limits variation to two degrees of freedom. This is accomplished by first taking the projection of the magnetic field onto the plane normal to the mean field, and then calculating the eigenvalues of the covariance matrix of the projected points. These eigenvalues represent the power in the variation along the maximum and minimum perpendicular variance directions. We measure the perpendicular variance isotropy of each interval, $\Psi$, which is defined as
\begin{equation}\label{eq:psi}
    \Psi = \frac{\lambda_2}{\lambda_1+\lambda_2},
\end{equation}
where $\lambda_i$ are the eigenvalues. Thus, when $\Psi = 0$, variation in the perpendicular plane is in only one direction, and when $\Psi = 0.5$, variation is equally distributed along two axes. This process is illustrated for two intervals shown in Figure \ref{fig:hodograms}.


To calculate the scaling behavior of the magnetic field, the second-order structure function, $\delta \mathbf{B}^2(\tau)$ is used: 
\[
\delta \mathbf{B}^2(\tau) = \langle|\mathbf{B}(t+\tau) - \mathbf{B}(t)|^2\rangle.
\]
The structure function is calculated over the inertial range, using lags $20s < \tau < 180s$, averaged over the whole interval. It is then fitted versus $\tau$ in log-log space using a least-squares linear fit; the power-law scaling obtained is $\alpha_B$. For an asymptotically long inertial range, this is related to the magnetic spectral index, $\gamma$, by
     \[
        \alpha_B = -(1+\gamma).
    \]

\section{Results} \label{sec:results}
    \begin{figure*}[ht!]
    \plotone{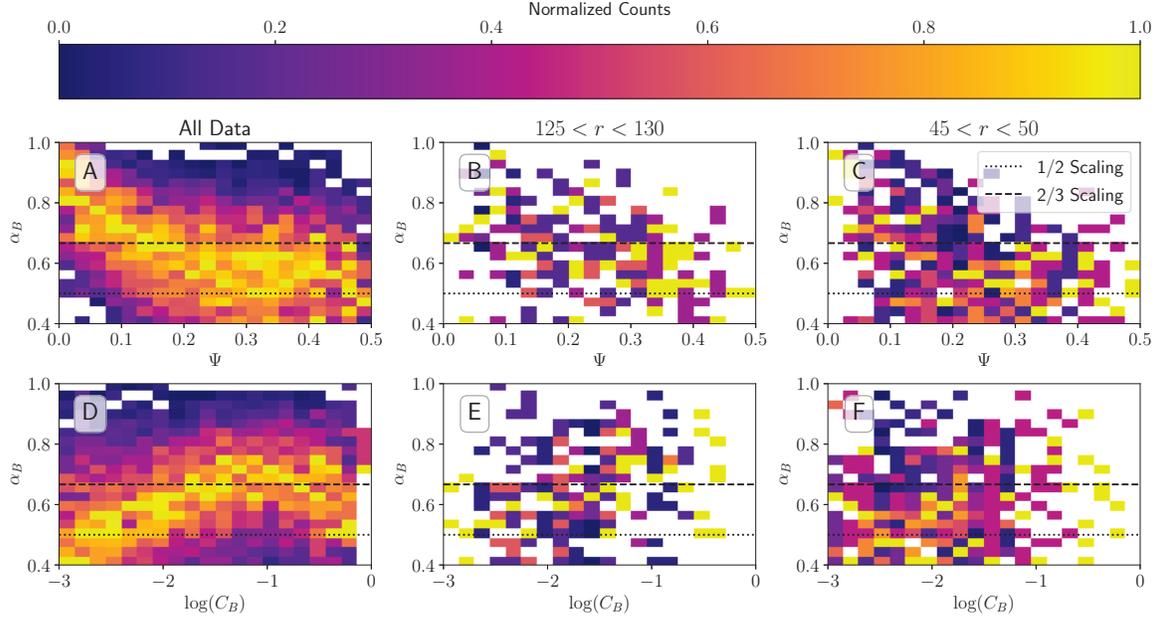}
    \caption{Panels A, B, and C show joint probability distributions of $\Psi$ and $\alpha_B$, with data column-normalized. Panels E, F, and G likewise show column-normalized joint probabilities of $\log(C_B)$ and $\alpha_B$. Panels A and E show all studied data, while panels B and F show data only in a  5 $R_\odot$ bin centered around 127.5$R_\odot$. Panels C and G similarly show a 5 $R_\odot$ at 47.5 $R_\odot$. The dashed black line in each panel marks 2/3 scaling, while the dotted black line marks 1/2 scaling.
    \label{fig:psiPhibyAlph}}
    \end{figure*}

    \subsection{Global Statistics}
    
    Categorization by $C_B$ and $\Psi$ as described above yields a peaked distribution around approximately arc-polarized structures ($\Psi = 0.165$; $\log(C_B) = -1.39$), as seen in Figure \ref{fig:3dminiatures}. The compressibility was found to be uniformly small, which is consistent with past results, and the assumption of an Alfvénic solar wind. The mean value of $C_B$ was found to be 0.067, representing that the compressive component was on average 26\% of the amplitude of the trace fluctuations. However, the quartiles of $C_B$ were 0.013, 0.034, and  0.081 respectively, reflecting a long tailed distribution with most observations (70\%) lower than the mean. Because $C_B << 1$ almost everywhere, variations in $C_B$ are henceforth analyzed through $\log(C_B)$, shown in Figure \ref{fig:3dminiatures}(B). 
    
    Perpendicular fluctuations were also most often mainly along one main axis, with a mean $\Psi$ of 0.22, representing that 88$\%$ of the variation power was along the maximum perpendicular variance direction.  Although the distribution was skewed left as seen in Figure \ref{fig:3dminiatures}(C), there were significant populations of observations exhibiting the full range of $\Psi$, with quartiles 0.13,  0.21,  and 0.30.  1.7$\%$ of observations had $\Psi > 0.45$, representing that the eigenvalue power along the minimum and maximum variance directions were within $5\%$ of the total eigenvalue power of each other.
    
    The distribution of $\alpha_B$ shows good agreement with previous measurements \citep{podesta2010kinetic, bowen2018impact, chen2013residual, chen2020evolution}. Most data-points were collected far from the sun due to the elliptical orbit of the spacecraft, with a median radial distance of 143$R_\odot$, and the mean observed $\alpha_B=0.65\approx 2/3$, as we expect \citep{podesta2010kinetic, chen2013residual, wicks2013correlations, roberts2010evolution, horbury2008anisotropic}. If data when the radial distance of the spacecraft from the sun $R > 200R_\odot$ is selected, a slightly steeper mean value of $\alpha_B = 0.68$ is observed. When data from $R<30R_\odot$ is selected, we observe a shallow $\alpha_B = 0.49$ mean, in close agreement with previous estimates of 1/2 \citep{chen2020evolution, sioulas2023magnetic}.
    
    Figure \ref{fig:3dminiatures}(A) also shows that the scaling behavior is a function of both $\Psi$ and the magnetic compressibility. An $\alpha_B \approx 1/2$ structure function scaling, matching analytical models of the solar wind \citep{chandran2015intermittency, mallet2017statistical, Boldyrev2006Spectrum}, is visible at low compressibility ($\log(C_B)<-2$) and when $\Psi > 0.25$. They match the characteristics of spherically polarized Alfvén waves, where the low magnetic compressibility shows that the constant magnitude condition is met. 
    
    The strongly one-dimensional structures we observe show very steep scaling (mean $\alpha_B = 0.80$ when $\Psi < 0.05$), consistent with observations of strong discontinuities, which have $\alpha_B = 1$ \citep{li2011effect, borovsky2010contribution}. These discontinuities would show strong variance along one axis, causing a low $\Psi$. The steepening structure functions with decreasing $\Psi$ suggests that the low $\Psi$ populations are increasingly influenced by discontinuities, which increase the mean $\alpha_B$.
    The generally reported 1AU scaling, 2/3, is observed in those observations which are neither entirely one dimensional nor extremely low compressibility. The steeper structure functions of the more compressive observations suggests that the less Alfvénic observations have a different characteristic scaling behavior. 
    
    \begin{figure*}[ht!]
    \plotone{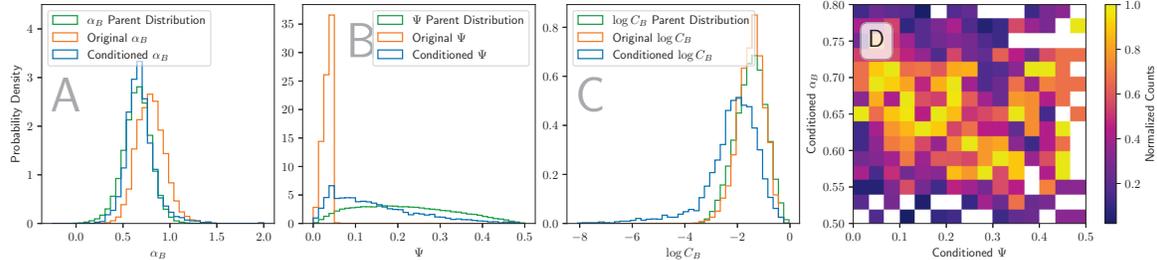}
    \caption{The figure shows the effect of the ``discontinuity conditioning" described on those extremely one dimensional observations where $\Psi < 0.05$. Panels A, B, and C show the effect on $\alpha_B$, $\Psi$, and $\log C_B$ respectively, where orange shows unconditioned values for $\Psi<0.05$, blue shows the conditioned distribution, and green shows the distribution for all intervals across the whole range of $\Psi$ (the ``parent distribution"). Panel D shows a column-normalized joint probability between the conditioned $\Psi$ and the conditioned $\alpha$ for the one dimensional observations.
    \label{fig:conditionedDists}}
    \end{figure*}
    
    \subsection{Correlation with Scaling Behavior}
    Correlation between $\Psi$ and $\alpha_B$ and between $\log(C_B)$ and $\alpha_B$ are further illustrated in Figure \ref{fig:psiPhibyAlph}. Figure \ref{fig:psiPhibyAlph}(A) shows the correlation between $\alpha_B$ and $\Psi$. This shows an interesting saturation behavior, where $\alpha_B$ does not significantly decrease for $\Psi>0.25$. This suggests that the correlation with $\Psi$ may only be a product of discontinuities, which appear for small $\Psi$, and that sufficiently isotropic perpendicular fluctuations produce the same scaling exponent. There is also no $\Psi$ for which $\alpha_B$ is distributed around the 1/2 scaling observed in the pristine solar wind, suggesting that while discontinuities, or other mechanisms resulting in small $\Psi$, may steepen the spectra beyond an $\alpha_B = 2/3$ scaling, that these processes do not determine the evolution from 1/2 to 2/3 spectral scalings observed in the solar wind. The spectral index and the magnetic compressibility are known to depend on the solar distance \citep{chen2020evolution, sioulas2023magnetic}, and we will show in Figure \ref{fig:radialTrends}(A) that $\Psi$ does as well. We therefore consider data constrained to two narrow ranges of radial distances, shown in Figure \ref{fig:psiPhibyAlph}(B,C). The trends discussed above were consistent between different radii, although $\alpha_B$ was generally lower for closer radii, as expected.
    
    However, we see a more complete range of scaling behavior when including analysis of the magnetic compressibility. Figure \ref{fig:psiPhibyAlph}(D) shows the correlation between the log of the magnetic compressibility and $\alpha_B$. Extremely incompressible intervals show an $\alpha_B = 1/2$ spectral index. Higher compressibility trends towards a steeper spectral index with a maximum average of about $\alpha_B = 2/3$, until $\log C_B = -0.25$, at which point we see a slight negative correlation. Observations of $\log C_B > -0.25$ occur very rarely (354 times -- 0.3\% of observations), so this apparent negative trend may simply be scatter. Figure \ref{fig:psiPhibyAlph}(E,F) show the same correlation with data constrained to two narrow ranges of radial distances, confirming that the correlation between $C_B$ and $\alpha_B$ is not a function of radial distance.
    
    \subsection{Impact of Discontinuities}
    To examine whether the trend towards steeper scaling for more one dimensional variation is due to changing distribution of discontinuities, a system of ``discontinuity conditioning" employed on the very one dimensional intervals ($\Psi < 0.05$) to discern what characteristics they would have if no discontinuities were present. We employ non-overlapping increments, 
    \[
         \mathbf{\delta B}(n, \tau) = \mathbf{B}((n+1)\tau) - \mathbf{B}(n\tau).
    \]
    
    Note that we can re-construct $\mathbf{B}(n\tau)$ by summing all the preceding increments, i.e.,
    \[
     \mathbf{B}(n\tau) = \mathbf{B}(0) + \sum_{i=0}^{n-1}\mathbf{\delta B}(i, \tau).
    \]
    The distribution of small scale (20s) non-overlapping increments is examined. Outliers further than 3 standard deviations from the mean were considered to be discontinuities. This choice of threshold did not significantly affect the resulting distributions when it was between 2$\sigma$ and 4$\sigma$. These increments are then subtracted from all points with time greater than $n\tau$. The timeseries should then be continuous but otherwise unaltered. The structure function, $\Psi$, and $C_B$ are then recomputed using the conditioned timeseries.

     The conditioned distributions, which attempt to represent the background statistics of the 4867 intervals where $\Psi < 0.05$ in the absence of discontinuities are shown in Figure \ref{fig:conditionedDists}. We see in Figure \ref{fig:conditionedDists}(A)  that the distribution of $\alpha_B$ is shifted significantly shallower, returning reasonably closely to the distribution of all intervals (the ``parent distribution"). The mean shifts from 0.80 to 0.68, comparable to the 0.65 mean of the parent distribution. This shift suggests that steep scaling was dominated by the presence of discontinuities, consistent with \cite{li2011effect, borovsky2010contribution}, and that the ``background" scaling for low $\Psi$ values without a dominant discontinuity was not significantly a different than the ensemble of observations for all $\Psi$. Similarly, we see in Figure \ref{fig:conditionedDists}(B) that the conditioned $\Psi$ is also distributed dramatically differently. Almost the full range of $\Psi$ values are represented, although the conditioned distribution is slightly peaked towards low $\Psi$. Thus, we see that the discontinuities also dominated the observed variation, where strong one-dimensional variance was almost always due primarily to these discontinuities. The remaining peak at low $\Psi$ may indicate that there also exist one dimensional structures remaining in the conditioned distribution. It may also be an artifact of the arbitrary threshold chosen to define discontinuities. Figure \ref{fig:conditionedDists}(D) shows the column-normalized joint probability between the conditioned $\Psi$ and conditioned $\alpha_B$. It should be read analogously to Figure \ref{fig:psiPhibyAlph}(A). We see that with discontinuities removed, correlation is much less convincing, if present at all. This supports the hypothesis that the increase in $\alpha_B$ as $\Psi$ decreases (Figure \ref{fig:psiPhibyAlph}(A)) is primarily due to the presence of discontinuities.
     
     Figure \ref{fig:conditionedDists}(C) shows the change in the magnetic compressibility distribution. Observations are shown to become much less compressive, suggesting that the compressive fluctuations were dominated by these discontinuities. However, the distribution is lowered significantly past the end of the parent distribution. Given that very incompressible non-discontinuity structures are certainly observed, it is unlikely that this distribution is a realistic representation of the compressibility of the continuous parts of these observations -- this distribution change most likely signals that the conditioning scheme does not preserve the compressive variation well. We avoid more detailed analysis of these ``conditioned" sets because significant uncertainty remains about the effects of this conditioning scheme. However, this result is a compelling confirmation that the one dimensional fluctuations are dominated by a discontinuity which steepens the scaling.

    For each of the discontinuities identified by the method above, it is then of interest what type of discontinuity we observe. The discontinuity normals are computed with MVA(minimum and maximum variance analysis) applied to the 40 seconds surrounding each discontinuity \citep{sonnerup1998minimum}. To improve the accuracy of MVA and avoid the effects of wave activity near the discontinuties, we only consider cases where the ratio between the intermediate and minimum eigenvalues is greater than 10 \citep{knetter2004four}. We then define $|\Delta |\mathbf{B}||$ as the difference in the magnitude of the field between the 60 seconds upstream of the discontinuity and the 60 seconds downstream, and $|B_n|$ as the mean magnitude of the magnetic field normal to the discontinuity. Discontinuity types are defined based on \cite{neugebauer1984reexamination}'s classification of discontinuities as rotational, tangential, either, or neither, as listed in Table \ref{tab:discontinuities}. This classification has also been used to understand the switchbacks observed by PSP \citep{larosa2021switchbacks, akhavan2021discontinuity}. 
    \begin{deluxetable}{ccc}
    \label{tab:discontinuities}
    \tablecaption{Discontinuity Type Criteria}
    
    \tablenum{1}
    
    \tablehead{\colhead{Type} & \colhead{$B_n$ Condition } & \colhead{$\Delta|\mathbf{B}|$ Condition}} 
    
    \startdata
    Rotational & $|B_n|/|\mathbf{B}| \geq 0.4$ & $|\Delta|\mathbf{B}||/|\mathbf{B}| < 0.2$ \\
    Tangential & $|B_n|/|\mathbf{B}| < 0.4$ & $|\Delta|\mathbf{B}||/|\mathbf{B}| \geq 0.2$  \\
    Either & $|B_n|/|\mathbf{B}| < 0.4$ & $|\Delta|\mathbf{B}||/|\mathbf{B}| < 0.2$ \\
    Neither & $|B_n|/|\mathbf{B}| \geq 0.4$ & $|\Delta|\mathbf{B}||/|\mathbf{B}| \geq 0.2$ \\
    \enddata
    
    \end{deluxetable}
    \\  

    Categorizing the discontinuities thus, we find 35.9\% rotational discontinuities, 9.0\% tangential discontinuities, 55.0\% either, and 0.1\% neither. The strong discontinuities we observe are on average at $154 R_\odot$, and these proportions are similar to previous results at 1AU (see \cite{neugebauer2006comment}) -- past results find between 10 and 15 percent tangential discontinuities, between 0 and 5 percent neither, and the rest are split between rotational discontinuities or ``either". The high proportion of ``either" designations we observe may be due to the low level of compressibility we see for the whole population, which requires that $|\Delta|\mathbf{B}||/|\mathbf{B}|$ is small. We see that discontinuities we select do not have significantly different statistics than those selected in past studies.

    \subsection{Radial Evolution}
    \begin{figure*}[ht!]
    \plotone{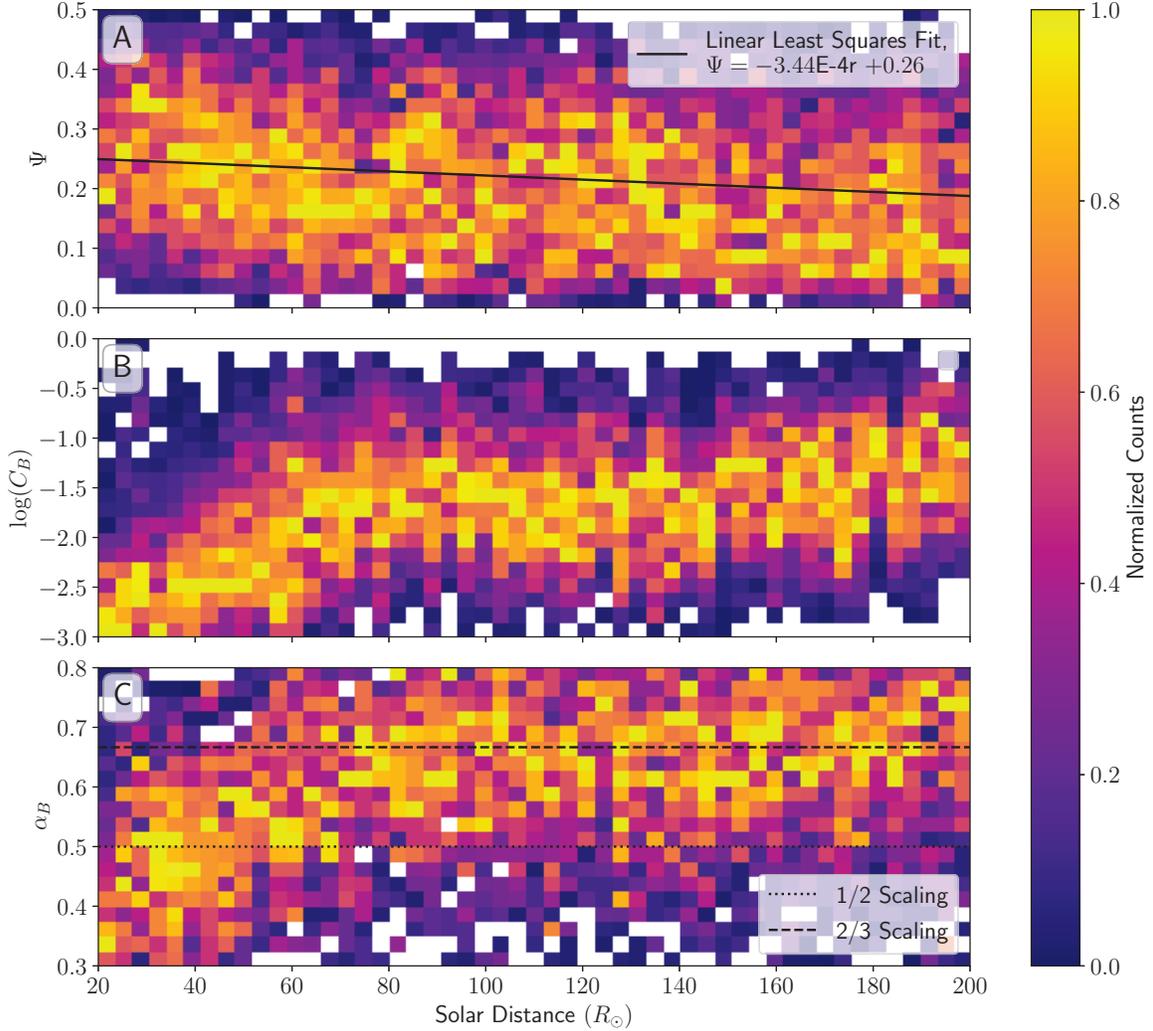}
    \caption{Panel A shows the evolution of $\Psi$ by solar distance as a column-normalized joint probability. A linear least squares fit is shown as a solid black line, $\Psi = -3.99 \times10^{-4}r + 0.27$. Panel B shows a column-normalized joint probability between the solar distance and $\log(C_B)$. Panel C shows a column-normalized joint probability between the solar distance and $\alpha_B$, with dashed and dotted black lines indicating respectively a 2/3 and 1/2 scaling. 
    \label{fig:radialTrends}}
    \end{figure*}

    The spectral index is a function of radius \citep{chen2020evolution, sioulas2023magnetic}, and this may be fundamentally linked to evolving fluctuation geometry. Figure \ref{fig:radialTrends}(A) shows that fluctuations become significantly less three dimensional as the solar wind propagates out from the Sun. A linear least-squares fit of all points yields the dependence $\Psi = (-3.99\times 10^{-4})r + 0.27$. More one dimensional fluctuation could be the result of increasing numbers of discontinuities generated by nonlinear interactions.
    Figure \ref{fig:radialTrends}(B) shows the radial trend of the magnetic compressibility. The wind becomes significantly more compressive outside 60$R_\odot$, then maintains a more constant value. \cite{chen2020evolution} reported the evolution of the magnetic compressibility as $C_B \propto r^{1.68\pm0.23}$, with significant scatter. Although this estimate covers the same range of values as this observation and has the same concavity, our result is not consistent with a power-law scaling. The dependence we do observe is notable in its similarity with that of the scaling exponent in Figure \ref{fig:radialTrends}(C). The major steepening of $\alpha_B$ we see in this range of radii happens between 20 $R_\odot$ and 60$R_\odot$, where we can see in Figure \ref{fig:radialTrends}(A) there are very few large discontinuities present. However, the relative size of the compressive fluctuations grow significantly. Then, the increase in $\alpha_B$ in this range may actually be conditioned on the magnetic compressibility.
    
\section{Conclusions}  \label{sec:conculsion}
Analytic models for the scaling behavior of three dimensional, anisotropic Alfvénic turbulence predict an $\alpha_B = 1/2$ structure function scaling \citep{chandran2015intermittency, mallet2017statistical, Boldyrev2006Spectrum}. This work also recovers an approximately 1/2 scaling exclusively when the magnetic compressibility (Eq. \ref{eq:CB}), $C_B$, is less than 0.01, and when the perpendicular fluctuations vary in two dimensions, with perpendicular variance isotropy (Eq. \ref{eq:psi}), $\Psi$,  greater than 0.25. These statistics match descriptions of spherically polarized Alfvén waves. The structure function scaling is found to increase significantly to $\alpha_B \approx 1$ for structures with variation in one extremely dominant direction ($\Psi<0.05$) (Figure \ref{fig:3dminiatures}). This steep scaling matches descriptions of the $\alpha_B = 1$ scaling of discontinuities \citep{li2011effect, borovsky2010contribution}. Upon numerically removing discontinuities from these low-$\Psi$ intervals, we find that the conditioned datasets have an underlying scaling behavior similar to the parent distribution, confirming that this steep scaling can be attributed to the presence of a large discontinuity (Figure \ref{fig:conditionedDists}). The observed one dimensional variation itself is also found by this scheme to be mainly due to these discontinuities. We categorize the discontinuities we find, which are on average at $154 R_\odot$, as $35.9\%$ rotational, $9.0\%$ tangential, $55.0\%$ either, and $0.1\%$ neither. These proportions are similar to previous results at 1AU \citep{neugebauer2006comment}. Thus, the strong, steeply scaling, one dimensional variation we observe matches previous descriptions of discontinuities in the solar wind.

Although very steep ($\alpha_B > 2/3$) scaling is associated with low $\Psi$, there is no $\Psi$ for which $\alpha_B$ is on average 1/2 (it is at lowest about 0.6), which suggests that distribution of discontinuities cannot fully explain the magnetic scaling exponent behavior, because the absence of discontinuities does not result in the $\alpha_B = 1/2$ scaling observed closer to the Sun (Figure \ref{fig:psiPhibyAlph}(A)). In addition to the effect of discontinuities, the structure function scaling is found to be conditioned by the magnetic compressibility. Incompressible fluctuations with $C_B < 0.01$ are found to have an $\alpha_B \lesssim 1/2$, with more compressible structures associated with $\alpha_B \approx 2/3$ (Figure \ref{fig:psiPhibyAlph}(D)). This steeper scaling for compressive structures may be due to an independent parallel component of the magnetic field with steeper scaling, which is consistent with the result of \cite{chapman2007quantifying} for the velocity fluctuations. The compressive fluctuations could be small-amplitude tangential discontinuities, which are known to have steep scaling \citep{tu1995mhd}. Alternatively, the compressive fluctuations are often attributed to the magnetosonic slow mode \citep{howes2012slow, verscharen2017kinetic, klein2012using}, which is coupled to the Alfvén mode by the parametric decay instability \citep{tenerani2013parametric, derby1978modulational, jayanti1993parametric}. Nonlinear energy exchange between the slow and Alfvén modes may then affect the scaling behavior. To summarise, our result suggests that a $\alpha = 1/2$ scaling is only possible when there are no compressive fluctuations present, and that an $\alpha = 1$ scaling is only observed in the presence of strong discontinuities.

Many previous authors have also observed that $\alpha_B$ increases from 1/2 in the inner heliosphere to 2/3 outside of about 60 $R_\odot$ \citep{podesta2010kinetic, chen2020evolution, wicks2013correlations, roberts2010evolution}. We observe that $\Psi$ is also correlated with solar distance, with $\Psi$ decreasing approximately linearly as the solar wind streams out from the Sun (Figure \ref{fig:radialTrends}). This may be a function of increasing importance of discontinuities. As mentioned above, the behaviour of $\Psi$ is strongly controlled by the presence or absence of discontinuities. We may also be seeing some process which progressively destroys three-dimensional, spherically-polarized Alfvén waves, for example reflections or the parametric decay instability. 

The magnetic compressibility also monotonically increases with radius. The very low $C_B$ we observe at small radii suggests that the primordial state of fluctuations in the corona is that of three-dimensional, spherically polarized Alfvén waves. The steepening of $\alpha_B$ which we see between 10 and 60 $R_\odot$ coincides with a large increase in $C_B$. Given that the corresponding measurements of $\Psi$ show that large discontinuities were rare in this range, the increase in compressibility may be a major factor in the increase of $\alpha_B$. The mechanism for the growth in the compressive fluctuations depends on their nature. The increase in compressibility may be an increased distribution of tangential discontinuities, which could be increasingly generated from pressure balance structures. If the compressive fluctuations are dominated by the slow mode, we may be seeing an \textit{in-situ} generation mechanism, e.g. the Alfvén wave parametric decay instability. The compressibility is intimately connected with the Alfvénicity and may serve as a proxy for the cross-helicity, which has previously been shown to be related to the spectral index \citep{sioulas2023magnetic, bowen2018impact, podesta2010scale, chen2013residual}. So, we can see the increase in compressibility as a breakdown of an Alfvénic equilibrium state, which coincides with spectral steepening. A study of the solar wind velocity fluctuations would lend insight into the relationship between $C_B$, the cross helicity, and the residual energy, and would reveal if the velocity spectrum is similarly dependent on the compressibility.


\begin{acknowledgments}
C. Dunn is supported by NASA PSP-GI Grant No. 80NSSC21K1771 as well as by PSP FIELDS funding trough NASA Contract No. NNN06AA01C.
\end{acknowledgments}

%


\software{numpy \citep{harris2020array}, scipy \citep{2020SciPy-NMeth}, matplotlib \citep{Hunter:2007}, pandas \citep{reback2020pandas}, sunpy \citep{sunpy_community2020}, astropy \citep{astropy:2022}
}



\bibliographystyle{aasjournal}



\end{document}